\documentclass[pra,twocolumn,showpacs,preprintnumbers,amsmath,amssymb]{revtex4}
\usepackage{epsfig}
\usepackage{color}
\usepackage{graphicx}
\usepackage{dcolumn}
\usepackage{bm}
\newcommand{\bra}[1]{\langle#1|}
\newcommand{\ket}[1]{|#1\rangle}

\providecommand{\openone}{\leavevmode\hbox{\small1\kern-3.8pt\normalsize1}}

\begin{document}
\title{Revival of quantum correlations without system-environment back-action}
\author{R. Lo Franco,$^{1,2}$ B. Bellomo,$^1$ E. Andersson,$^3$ and G. Compagno$^1$}
\affiliation{$^1$CNISM and Dipartimento di Fisica, Universit\`a di Palermo - via Archirafi 36, 90123 Palermo, Italy\\
$^2$Centro Siciliano di Fisica Nucleare e di Struttura della Materia (CSFNSM) and Dipartimento di Fisica e Astronomia, Universit\`a di Catania - Viale A. Doria 6, 95125 Catania, Italy\\
$^3$SUPA, Department of Physics, Heriot-Watt University - Edinburgh EH14 4AS, United Kingdom}

\date{\today}

\begin{abstract}
Revivals of quantum correlations have often been explained in terms of back-action on quantum systems by their quantum environment(s). Here we consider a system of two independently evolving qubits, each locally interacting with a classical random external field. The environments of the qubits are also independent, and there is no back-action on the qubits. Nevertheless, entanglement, quantum discord and classical correlations between the two qubits may revive in this model. We explain the revivals in terms of correlations in a classical-quantum state of the environments and the qubits. Although classical states cannot store entanglement on their own, they can play a role in storing and reviving entanglement. It is important to know how the absence of back-action, or modelling an environment as classical, affects the kind of system time evolutions one is able to describe. We find a class of global time evolutions where back-action is absent and for which there is no loss of generality in modelling the environment as classical. Finally, we show that the revivals can be connected with the increase of a parameter used to quantify non-Markovianity of the single-qubit dynamics.
\end{abstract}


\maketitle

\section{Introduction}

Understanding the dynamics of correlations in quantum systems is essential for quantum information and quantum computation \cite{nielsenchuang,eberly2009Science}, and this has been thoroughly investigated for quantum systems in quantum environments. In this paper, we look at the behavior of correlations for the case when the environment is modelled as classical, and when the environment is unaffected by the system dynamics. This implies that there can be no back-action on the system via the environment.

Sometimes only parts of the total correlations may be relevant. In particular, one can characterize, via appropriate quantifiers, the quantum and classical parts of the correlations \cite{Zurek2001PRL,vedral2001JPA}. The quantum part of the correlations, quantified by quantum discord, includes not only entanglement but also correlations that may occur in separable states. For bipartite open quantum systems in quantum environments, entanglement may display phenomena such as sudden death, revivals and trapping \cite{yu2004PRL,bellomo2007PRL,bellomo2008trapping}. Revivals of entanglement, after finite time periods when it completely disappears, can be expected when either direct two-qubit interactions \cite{horodecki2001PRA,tanas2006} or indirect effective interaction  are present,  as in the case of two qubits in a common quantum reservoir \cite{mazzola2009PRA}. This is because interactions among quantum systems can create or destroy correlations, in particular, quantum correlations. Revivals may also occur for noninteracting qubits in independent quantum environments and have then been related to the non-Markovian nature of the environments \cite{bellomo2007PRL,bellomo2008PRA}. Recently, attention has been moved to the dynamics of correlations other than entanglement in the presence of either Markovian \cite{werlang2009PRA} and non-Markovian \cite{fanchini2010PRA,wang2010PRA} quantum environments. A peculiar aspect of the dynamics of these correlations is that even when total correlations behave in a regular way, both classical correlations and quantum discord may remain frozen for finite time intervals \cite{mazzola2010PRL}.

In the case of initially entangled noninteracting qubits in independent non-Markovian quantum reservoirs, entanglement revivals have been explained in terms of transfer of correlations back and forth from the two-qubit system to the various parts of the total system. This is due to the back-action via the environment on the system, which creates correlations
between qubits and environments and between the environments themselves. In this paper we will refer to this kind of correlations as correlations induced by back-action. In particular, correlations may build up between the two independent quantum reservoirs \cite{lopez2008PRL,bai2009PRA,Lopez2010PRA} and this phenomenon has been named sudden birth of entanglement in reservoirs.

On the other hand, there are features that classical models of quantum systems fail to capture. Similarly, one would expect that if the quantum environment of a quantum system is modelled as classical, for example a classical light field coupled to a quantized atom, then there might be qualities of the dynamics that this model is unable to describe. Intuitively, one might expect that since a classical environment should not be able to store quantum correlations in the same way as a quantum environment, the ability to describe revivals of quantum correlations might be affected. Similarly, if the environment has no back-action on the system, then it would seem that this could also affect the correlation dynamics. Nevertheless, revivals of quantum correlations can occur for random classical telegraph noise~\cite{zhou2010QIP}. It is clearly important to know exactly what kinds of behavior of correlation dynamics can be described when an environment is modelled as classical instead of quantum, or when back-action on the system is not present.

Existing work thus mostly concerns quantum systems in quantum environments, and has explained revivals in terms of back-action on the systems by their quantum environments, and in terms of quantum correlations involving the quantum environments. In this paper, we want to explain how and why revivals of correlations, including quantum correlations, can generally  occur also if the environment is classical, and when no back-action is present. We illustrate our discussion with a model characterized by the absence of correlations induced by back-action, where the element of randomness is introduced in a very simple way. The example we will consider is a pair of independent qubits driven by single classical field modes with random phase, where revivals of correlations do occur both for entanglement, for the so-called quantum discord, and for classical correlations.

We explain the revivals in terms of correlations in the classical-quantum state of the environments and qubits. We then discuss a class of global time evolutions for which back-action on the system is absent. This certainly occurs when the environment is unaffected by the system. We present a necessary condition for this to be the case. If the environment is initially in a classical state, this condition is also sufficient for it to be unaffected by the system. The state of a quantum environment may however change.
Interestingly, for this general class of time evolutions, modelling the environment as classical results in no loss of generality. In other words, any time evolution of the system that could be obtained by coupling it to a quantum environment can also be obtained by modelling the environment as classical. Finally, we associate the revivals with the increase of a parameter that quantifies non-Markovianity of the reduced dynamics of the qubit.

\section{Correlation quantifiers}
Given a state $\rho$, we adopt the definitions in \cite{Modi2010PRL} to quantify total correlations $T(\rho)$, quantum discord $D(\rho)$, entanglement $E(\rho)$ and classical correlations $C(\rho)$,
{\setlength\arraycolsep{1.5pt}\begin{eqnarray}\label{quantifiers}
T(\rho)&\equiv& S(\rho\|\pi_\rho)=S(\pi_\rho)-S(\rho),\nonumber\\
D(\rho)&\equiv& S(\rho\|\chi_\rho)=S(\chi_\rho)-S(\rho),\nonumber\\
E(\rho)&\equiv& S(\rho\|\sigma_\rho),\nonumber\\
C(\rho)&\equiv& S(\chi_\rho\|\pi_{\chi_\rho})=S(\pi_{\chi_\rho})-S(\chi_\rho),
\end{eqnarray}}where $S(\rho\|\sigma)\equiv-\mathrm{Tr}(\rho\log \sigma)-S(\rho)$ is the relative entropy, $S(\rho)\equiv-\mathrm{Tr}(\rho\log\rho)$ is the von Neumann entropy; $\pi_\rho$, $\chi_\rho$ and $\sigma_\rho$ are respectively the product state, the classical state and the separable state closest to $\rho$, while $\pi_{\chi_\rho}$ is the product state closest to $\chi_\rho$. These states are such that they minimize the corresponding relative entropies. Note that entanglement $E$ cannot be expressed, in general, as a difference of state entropies. The definition of discord $D(\rho)$ followed here and the original one \cite{Zurek2001PRL} in general do not coincide, but do so for Bell-diagonal states \cite{Modi2010PRL}. We observe that the maximization procedure involved in the original definition of quantum discord has been analytically solved only for certain classes of quantum states, such as ``X" states \cite{Ali2010PRA}. Our analysis will be focused on the class of Bell-diagonal states, which is a subclass of ``X" states, so that known analytical results \cite{Luo2008PRA,Modi2010PRL} apply.

For two independent quantum systems each interacting with its own local environment, so that neither direct nor mediated interactions between the two quantum systems exist, it is known that entanglement and total correlations  cannot exceed their initial values \cite{plenio2007QIC, divincenzo2002JMP}. Using the definition of $T(\rho)$ in Eq.~(\ref{quantifiers}), the definition of the product state closest to $\rho$ and the ``generalised H-theorem'' for dynamical maps $S(\Lambda\rho\|\Lambda\sigma)\leq S(\rho\|\sigma)$ \cite{alickibook}, this property can be straightforwardly generalised to the case of a $N$-partite system where each independent subsystem is locally affected by an arbitrary completely positive (CP) map. We note that in contrast to this, quantum discord can increase with respect to its initial value under local operations \cite{dakic2010PRL}.

\section{Model\label{model}}
We consider a pair of noninteracting qubits each locally coupled to a random external field, whose characteristics are unaffected by the qubit it is coupled to. This implies that back-action on the dynamics of the qubits is absent. The aim is to  give a clear example of how revivals of quantum correlations may occur even without back-action. Each environment is a classical field mode with amplitude fixed and equal for both qubits. The phase of each mode is not determined, and is equal either to zero or to $\pi$ with probability $p=1/2$ (the case when $p_1\neq 1/2$ shall be treated in Sec.~\ref{sec:p1different}). This model describes a special case of two qubits each subject to a phase noisy laser \cite{andersson2007JMO,cresser2010OptComm} but where the phase can take only two values and with the diffusion coefficient in the master equation equal to zero. This model has in common with the one considered in Ref.~\cite{zhou2010QIP} that the environments are unaffected by the system of interest. However, here the element of randomness is introduced in the phase which can assume only two possible values with assigned and fixed probabilities without switching between them. In Ref.~\cite{zhou2010QIP} qubits are subject to random telegraph noise, where instead it is the coupling that switches between two values during the evolution.

For the case we are considering, the dynamical map for the single qubit $S=A,B$ is of the random external fields type \cite{alickibook,horodecki2001PRA} and can be written as
\begin{equation}
\label{eq:singlemap}
\Lambda_S(t,0)\rho_S(0)=\frac{1}{2}\sum_{i=1}^2U_i^{S}(t)\rho_S(0)U_i^{S\dag}(t),
\end{equation}
where $U_i^{S}(t)=\mathrm{e}^{-\mathrm{i}H_it/\hbar}$ is the time evolution operator with $H_i=\mathrm{i}\hbar g(\sigma_+e^{-\mathrm{i}\phi_i}-\sigma_-e^{\mathrm{i}\phi_i})$ and the factor $1/2$ arises from the equal field phase probabilities of the model (more in general, there is a probability $p_i^S$ associated to $U_i^S$). Each Hamiltonian $H_i$ is expressed in the rotating frame at the qubit-field resonant frequency $\omega$. In the basis $\{\ket{1},\ket{0}\}$, the time evolution operators $U_i^{S}(t)$ have the matrix form
\begin{equation}\label{unitarymatrix}
U_i^{S}(t)=\left(
\begin{array}{cc}\cos(gt)&\mathrm{e}^{-\mathrm{i}\phi_i}\sin(gt)\\
-\mathrm{e}^{\mathrm{i}\phi_i}\sin(gt) & \cos(gt) \\\end{array}\right),
\end{equation}
where $i=1,2$ with $\phi_1=0$ and $\phi_2=\pi$.

\begin{figure}
\begin{center}
\includegraphics[width=0.44\textwidth]{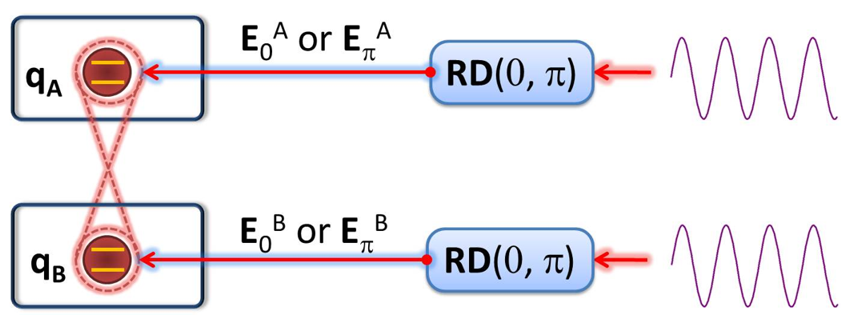}
\caption{\label{fig:system}\footnotesize (Color online) Sketch of the physical system. The two qubits are initially entangled.  Each random dephaser (RD) is such that the field phase at the location of each qubit is either zero or $\pi$ with equal probability, $p=1/2$, corresponding to either $\mathbf{E}^S_{0}$ or $\mathbf{E}^S_{\pi}$ ($S=A,B$).}
\end{center}
\end{figure}
This simple model is depicted in in Fig.~\ref{fig:system}, where the two local environments are realized by two single classical field modes of the same frequency and amplitude, but each passing across a random dephaser such that the phase of the field at the location of each qubit is either zero or $\pi$, with probability $p=1/2$. The interaction between each qubit and its local field mode is assumed to be strong enough so that, for sufficiently long times, the dissipation effects of the vacuum radiation modes on the qubit dynamics can be neglected. The dynamics is cyclic, as evinced from the fact that the unitary matrix of Eq.~(\ref{unitarymatrix}) (for each $i=1,2$) becomes the identity for times $t=k\pi/g$ ($k=0,1,2,\ldots$). The same initial state is thus retrieved at these times. If one takes into account dissipation effects on the system, the dynamics would not be cyclic anymore, but it is expected that for dissipation weak enough the qualitative behavior of the dynamics would remain the same until a characteristic time. This could be feasible by considering as qubits two atoms placed in separated zero temperature cavities. Each atom is subject to a resonant interaction with a laser with a random phase but out of resonance with respect to the characteristic cavity frequencies in order to slow down spontaneous emission.

The global dynamical map $\Lambda$ applied to an initial state $\rho(0)$ of the bipartite system,  $\rho(t)\equiv\Lambda(t,0)\rho(0)$, is again of the random external fields type, that is,
\begin{equation}\label{globalrandomfieldmap}
\rho(t)=\frac{1}{4}\sum_{i,j=1}^2U_i^{A}(t)U_j^{B}(t)\rho(0)U_i^{A\dag}(t)U_j^{B\dag}(t).
\end{equation}
We now show that the map $\Lambda$ moves inside the class of Bell-diagonal states (or states with maximally mixed marginals \cite{Luo2008PRA}). We first introduce the notation $\ket{1_\pm}\equiv(\ket{01}\pm\ket{10})/\sqrt{2}$ for the one-excitation Bell states and $\ket{2_\pm}\equiv(\ket{00}\pm\ket{11})/\sqrt{2}$ for the two-excitation Bell states. By using Eqns.~(\ref{unitarymatrix}) and (\ref{globalrandomfieldmap}), we find that the map acts on single Bell states as
\begin{equation}
\Lambda(t,0)\ket{i_\pm}\bra{i_\pm}=[1-f(t)]\ket{i_\pm}\bra{i_\pm}+f(t)\ket{i'_\mp}\bra{i'_\mp},
\end{equation}
where $i,i'=1,2$ with $i\neq i'$ and $f(t)=\sin^2(2gt)/2$, so that mixtures of the states $\ket{1_+}$ ($\ket{1_-}$) and $\ket{2_-}$ ($\ket{2_+}$) are mapped onto different mixtures of the same states. It immediately follows that the map $\Lambda$ moves inside the class of Bell-diagonal states, with a generic initial Bell-diagonal state written as
\begin{equation}\label{initialBelldiagonalstate}
\rho(0)=\sum_{i,s}\lambda_i^s(0)\ket{i_s}\bra{i_s},\quad (i=1,2;\ s=\pm).
\end{equation}

It is worth noting that the diagonal Bell states in Eq.~(\ref{initialBelldiagonalstate}) include the well-known Werner states \cite{bellomo2008PRA} and that $\rho(0)$ is entangled if the largest $\lambda_i^s(0)$ is greater than $1/2$ \cite{Modi2010PRL}. The action of the map (\ref{globalrandomfieldmap}) on $\rho(0)$ thus gives another Bell-diagonal state $\rho(t)$ with time-dependent coefficients
\begin{equation}\label{evolvedlambdacoefficients}
\lambda_i^\pm(t)=\lambda_i^\pm(0)[1-f(t)]+\lambda_{i'}^\mp(0)f(t) \quad(i\neq i').
\end{equation}
It can be shown that, for all $t\geq0$, the largest $\lambda_i^s(t)$ is always the same, while, depending on the initial values of the coefficients, there are switching times after which the second largest $\lambda_i^s(t)$ changes. Moreover, since the function $f(t)$ is periodic, there will be repeated switching times during the dynamics. The restriction of the dynamical map of Eq.~(\ref{globalrandomfieldmap}) to the class of Bell-diagonal states and the order between the $\lambda_i^s(t)$ is crucial to construct the classical, separable and product states closest to $\rho(t)$ and thus to analytically calculate the correlation quantifiers in Eq.~(\ref{quantifiers}) \cite{Modi2010PRL}. In particular, the closest product states do not evolve ($\pi_t=\Lambda\pi_0=\pi_0$) and are all equal to the normalized $4\times4$ identity $\mathbb{I}/4$, while the classical state closest to $\rho(t)$ is equal to the evolved classical state closest to $\rho(0)$, that is $\chi_{\rho(t)}\equiv\chi_{\Lambda\rho(0)}=\Lambda\chi_{\rho(0)}$. These properties permit us, by using the generalised H-theorem, to show that
{\setlength\arraycolsep{1.5pt}\begin{eqnarray}\label{DCinequality}
D[\rho(t)]&\equiv& S(\rho(t)\|\chi_{\rho(t)})\leq S(\rho(0)\|\chi_{\rho(0)})\equiv D[\rho(0)],\\
C[\rho(t)]&\equiv& S(\chi_{\rho(t)}\|\pi_{\chi_{\rho(t)}})
\leq S(\rho(0)\|\pi_{\rho(0)})\equiv C[\rho(0)].\nonumber
\end{eqnarray}}Therefore, under the map of Eq.~(\ref{globalrandomfieldmap}) not only entanglement but also discord and classical correlations cannot exceed their initial values during the dynamics.

\section{Dynamics of correlations for equal field phase probabilities}
\begin{figure}
\begin{center}
\includegraphics[width=0.40\textwidth]{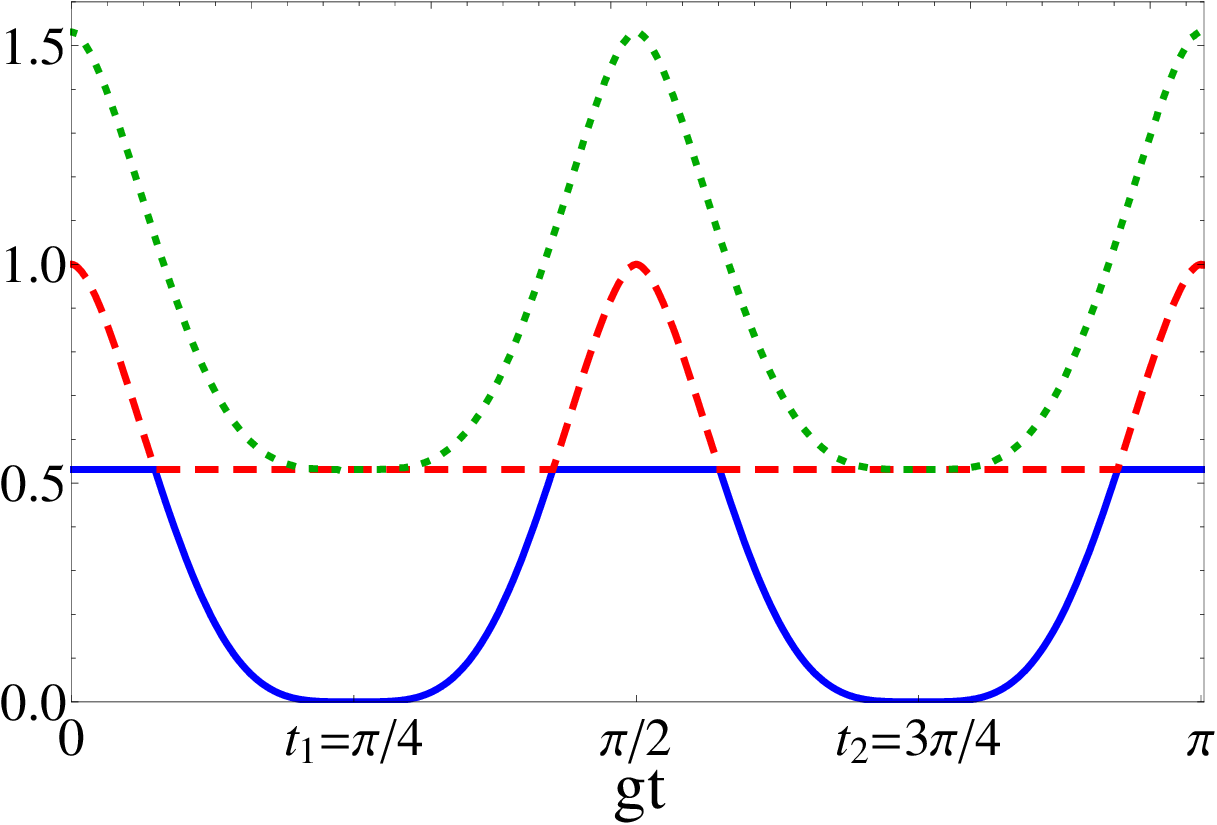}
\caption{\label{fig:correlations}\footnotesize (Color online) Dynamics of discord $D$ (blue solid line), classical correlations $C$ (red dashed line) and total correlations $T$ (green dotted line) for an initial Bell-diagonal state with $\lambda_1^+(0)=0.9$, $\lambda_1^-(0)=0.1$ and $\lambda_2^+(0)=\lambda_2^-(0)=0$. Total correlations oscillate regularly, while transitions between decay and rise of quantum and classical correlations occur periodically.}
\end{center}
\end{figure}

We are now able to analyze the evolution of the relevant quantifiers of correlations under the map of Eq.~(\ref{globalrandomfieldmap}) for different choices of $\rho(0)$. One finds a rich variety of behaviors of the quantifiers. In particular, Fig.~\ref{fig:correlations} shows the time evolutions of $T$, $D$ and $C$ starting from a Bell-diagonal state consisting of only the two Bell states $\ket{1_\pm}$, with $\lambda_1^+(0)=0.9$, $\lambda_1^-(0)=0.1$. This is a case for which switching times for the time-dependent coefficients $\lambda_2^-(t)$ and $\lambda_1^-(t)$ exist. In this case the dynamics of correlations is such that, while total correlations oscillate, either classical correlations or discord are frozen during alternating finite time periods. When discord is frozen, classical correlations are oscillating, and vice versa.  This kind of behavior has already been found in the case of two qubits locally subject to non-dissipative channels \cite{mazzola2010PRL}. This may thus be taken to signify that such  peculiar behavior of both quantum and classical correlations is a general result not connected to specific systems and interactions.

\begin{figure}
\begin{center}
\includegraphics[width=0.40\textwidth]{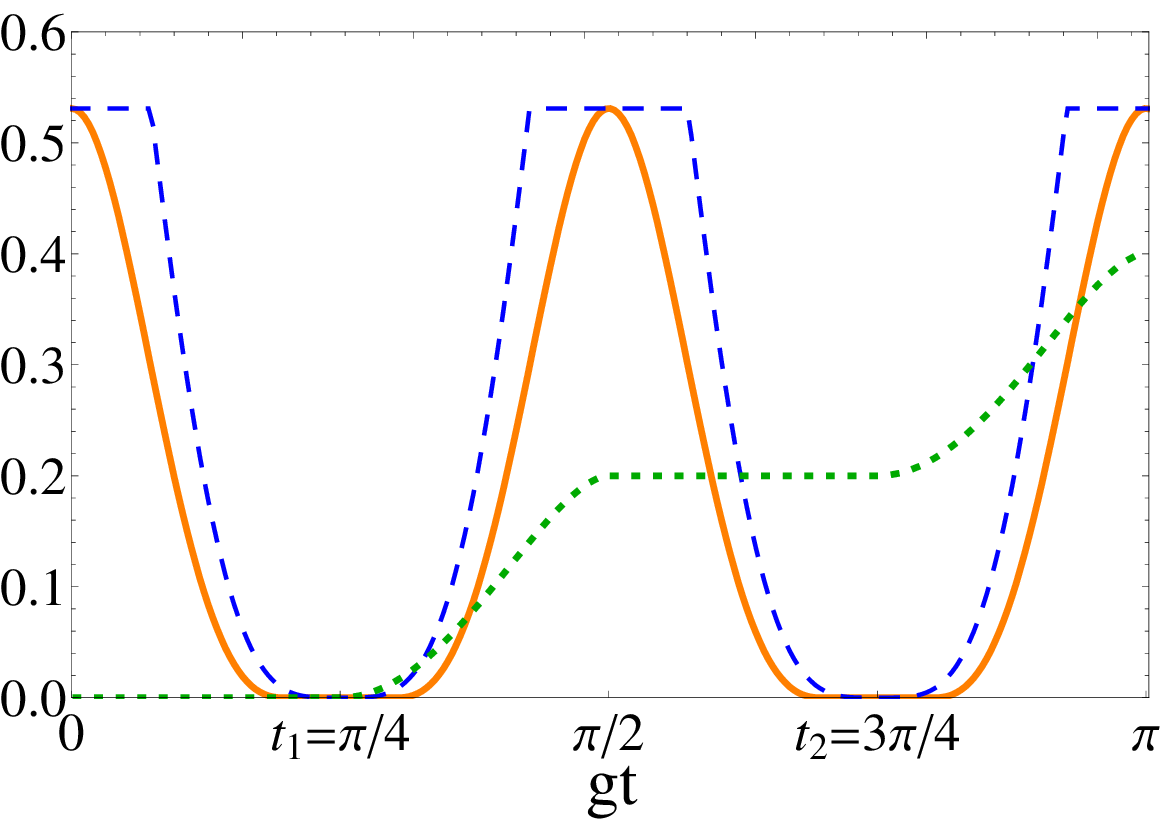}
\caption{\label{fig:entanglement}\footnotesize (Color online) Dynamics of entanglement $E$ (orange solid line) and discord $D$ (blue dashed line) for an initial Bell-diagonal state with the same coefficients as in Fig.~\ref{fig:correlations}. $D$ vanishes only at given times $t_n=(2n-1)\pi/(4g)$ ($n=1,2,\ldots$), while $E$ disappears for finite time intervals. The quantifier of non-Markovianity $\mathcal{I}^{E}(t)/10$ (green dotted line) is also plotted, showing its increase in correspondence with the revival of correlations.}
\end{center}
\end{figure}

Fig.~\ref{fig:entanglement} displays the dynamics of entanglement $E$ starting from the same initial Bell-diagonal state as before. Again, collapses and revivals of entanglement occur, as is known to happen for  quantum non-Markovian environments such as amplitude damping channels \cite{bellomo2007PRL}. In our case, the occurrence of revivals is clearly connected to the cyclical nature of the total dynamics. We note that in presence of small dissipation, the same qualitative behavior of correlations is expected, that is, partial revivals should still occur when the dissipation effects are weak enough. We also note that the occurrence of entanglement revivals is not in conflict with the fact that the two qubits are subject to local independent environments. Indeed, while entanglement must be constant if the local actions are unitary, in the general case for non-unitary dynamics (as in our model)  the only constraint (already mentioned above) is that the entanglement cannot exceed its initial value during the dynamics. This is not in conflict with the fact that, for example, entanglement goes to zero and then revives up to the initial value.

It should be pointed out that the above revivals of quantum correlations, as characterized by entanglement and discord, occur without transfer of correlations induced by back-action from system to environment. The absence of back-action implies that the initial two-qubit correlations cannot be ``temporarily stored" as correlations entirely within the environments. This differs from a common point of view where revivals are interpreted in terms of transfer of correlations induced by back-action which correlate the independent non-Markovian environments, which in turn again re-correlate the qubits \cite{lopez2008PRL,bai2009PRA,Lopez2010PRA}. In our example this mechanism is absent.

\section{Dynamics of correlations for different field phase probabilities}\label{sec:p1different}
The above results are obtained in the case when the phase of each field is zero or $\pi$ with equal probability distributions for both the subsystems ($p_1^S=p_2^S=1/2$, $S=A,B$). Extending the map of Eq.~(\ref{globalrandomfieldmap}) to the case when $p_1^S=1-p_2^S\neq1/2$ it becomes
\begin{equation}\label{globalmappipj}
\Lambda(t,0)\rho(0)=\sum_{i,j=1}^2p_i^Ap_j^BU_i^{A}(t)U_j^{B}(t)\rho(0)U_i^{A\dag}(t)U_j^{B\dag}(t).
\end{equation}
This map, applied to an arbitrary Bell-diagonal state of Eq.~(\ref{initialBelldiagonalstate}), in the standard computational basis $\mathcal{B}=\{\ket{1}\equiv\ket{11},\ket{2}\equiv\ket{10}, \ket{3}\equiv\ket{01}, \ket{4}\equiv\ket{00}\}$ gives the state at time $t$
\begin{equation}\label{densitymatrixp1}
\bar{\rho}(t)=\left(
\begin{array}{cccc}
  a(t) & b(t) & c(t) & d(t) \\
  b(t) & \frac{1}{2}-a(t) & e(t) & -c(t) \\
  c(t) & e(t) & \frac{1}{2}-a(t) & -b(t) \\
  d(t) & -c(t) & -b(t) & a(t) \\
\end{array}
\right),
\end{equation}
where
\begin{eqnarray}
a(t)&=&[1-\lambda_1^+(0)-\lambda_1^-(0)]/2-[F(\lambda)-G(\lambda)I(p)]f(t),\nonumber\\
b(t)&=&[F(\lambda)-p_1^AG(\lambda)-p_1^BL(\lambda)]\sin(4gt)/4,\nonumber\\
c(t)&=&[F(\lambda)-p_1^BG(\lambda)-p_1^AL(\lambda)]\sin(4gt)/4,\nonumber\\
d(t)&=&[F(\lambda)-I(p)L(\lambda)]f(t)-\lambda_2^-/2,\nonumber\\
e(t)&=&d(t)+[1-2\lambda_1^-(0)-\lambda_2^+(0)]/2,
\end{eqnarray}
with $F(\lambda)=1-2\lambda_1^+(0)-\lambda_1^-(0)-\lambda_2^+(0)/2$, $G(\lambda)=1-2\lambda_1^+(0)-\lambda_2^+(0)$, $L(\lambda)=1-2[\lambda_1^+(0)+\lambda_1^-(0)]$, $I(p)=p_1^A+p_1^B-2p_1^Ap_1^B$ and $f(t)=\sin^2(2gt)/2$, as before. 
The density matrix $\bar{\rho}(t)$ above is not Bell-diagonal in the basis $\mathcal{B}$ \cite{Luo2008PRA}, but it is straightforward to show that it has maximally mixed marginals, $\mathrm{Tr}_B\bar{\rho}(t)=\mathrm{Tr}_A\bar{\rho}(t)=\openone/2$, just like Bell-diagonal states. In fact, it is known that any two-qubit state with maximally mixed marginals always can be reduced, up to local unitary equivalence, to a state which is Bell-diagonal in some other basis \cite{rudolph2004JMP,Luo2008PRA}. Since local unitary operations do not affect the correlation quantifiers \cite{vedral2001JPA}, we can use the expressions of Eq.~(\ref{quantifiers}) to study the dynamics in terms of the eigenvalues of $\bar{\rho}(t)$, analogously to what was done above for the case $p_1^S=1/2$. Notice that, as a consequence, classical correlations can be obtained by $C=T-D$, a relation valid only for Bell-diagonal states \cite{Modi2010PRL}. 

\begin{figure}
\begin{center}
{\includegraphics[width=0.40\textwidth]{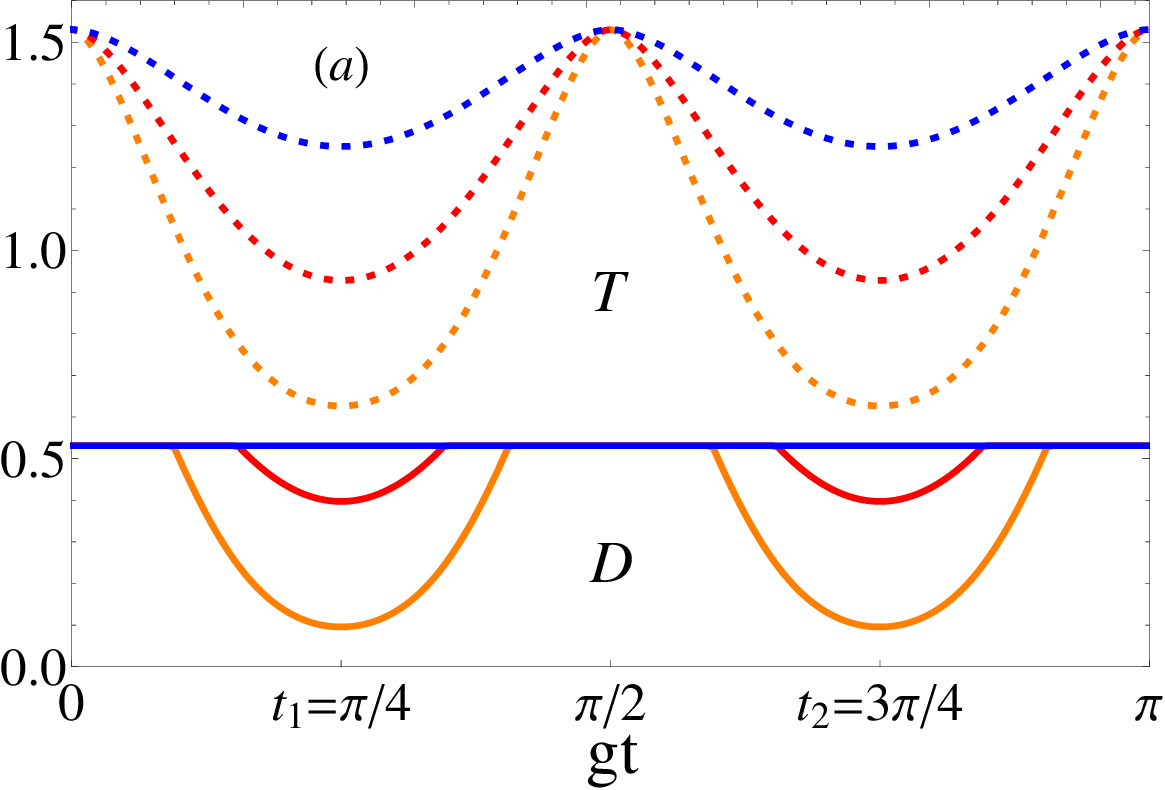}\vspace{0.5 cm}
\includegraphics[width=0.40\textwidth]{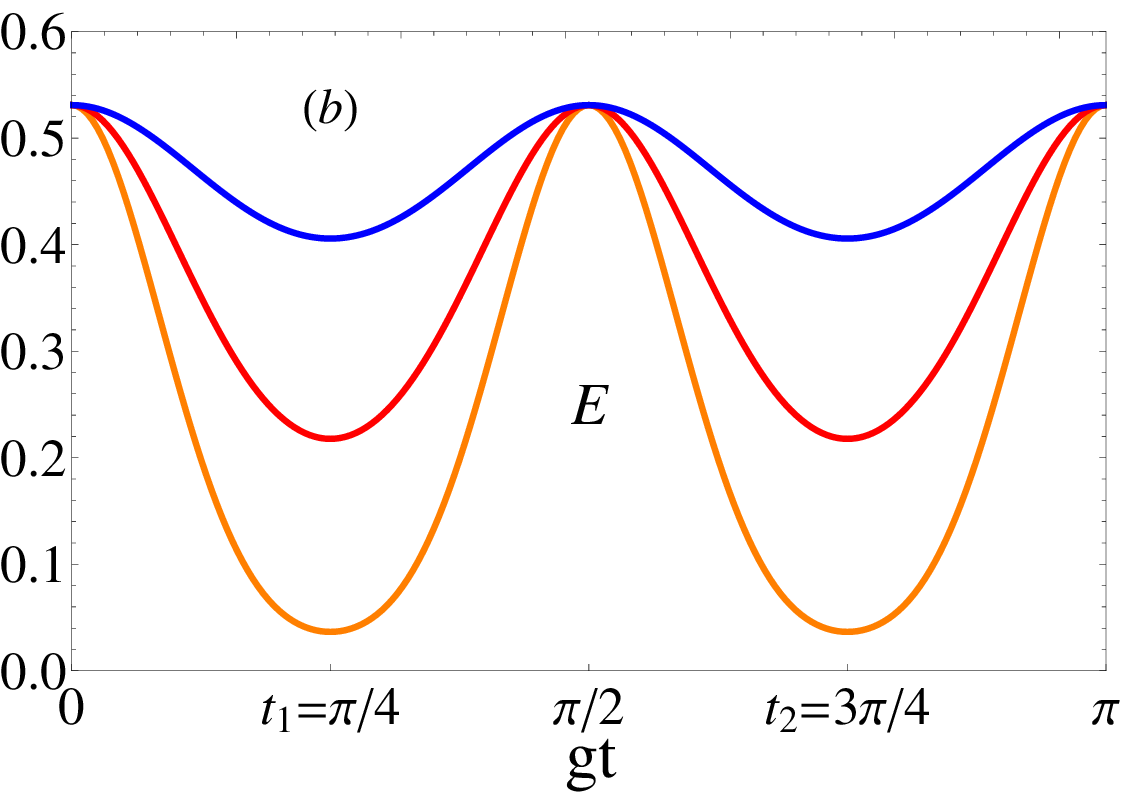}}
\caption{\label{fig:EDTp1}\footnotesize (Color online) Dynamics of correlation quantifiers $T$ (dashed lines, panel a), $D$ (solid lines, panel a) and $E$ (panel b) for an initial Bell-diagonal state with the same coefficients as in Fig.~\ref{fig:correlations} and for values of $p_1=1-p_2$ from bottom to top equal to 0.2 (orange line), 0.08 (red line), 0.025 (blue line). Discord $D$ is almost constant for $p_1=0.025$.}
\end{center}
\end{figure}

For the sake of simplicity we limit the present analysis to the case when $p_1^A=p_1^B=p_1$, because the other cases $p_1^A\neq p_1^B$ only result in quantitative but not qualitative differences in the time behavior of the correlation quantifiers. We are able to plot total correlations $T(\bar{\rho}(t))$, discord $D(\bar{\rho}(t))$ and entanglement $E(\bar{\rho}(t))$ as functions of $gt$ by choosing suitable values of the initial coefficients $\lambda_i^s(0)$ and diagonalizing the state $\bar{\rho}(t)$ for some values of $p_1$ at different values of $t$. In Fig.~\ref{fig:EDTp1} we in particular display the plots for the same initial Bell-diagonal state as was chosen for the case $p_1^S=1/2$ (that is, $\lambda_1^+(0)=1-\lambda_1^-(0)=0.9$ and $\lambda_2^\pm(0)=0$).
It is seen that the amplitudes of oscillation for all the correlation quantifiers tend to decrease when $p_1$ moves from $1/2$ (the behavior is symmetric with respect to $1/2$), with the quantifiers remaining closer to their initial values. In particular, quantum discord and entanglement do not vanish anymore. This behavior is expected, because changing $p_1$ cannot affect the presence or the frequency of oscillations, as is clear from the cyclic behavior of the dynamics due to Eqs.~(\ref{unitarymatrix}) and (\ref{globalmappipj}). Moreover, when the limiting case is considered where each local field becomes fully deterministic ($p_1^S=0,1$), the two-qubit evolution is a tensor product of local unitary evolutions on each qubit (see Eq.~(\ref{globalmappipj})), both when $p_1^A=p_1^B$ (fields in phase)  and when $p_1^A=1-p_1^B$ (fields with opposite phase). In this case all correlation quantifiers are, as known \cite{vedral2001JPA}, constant in time and equal to their initial value: the only possible behavior for correlation quantifiers when $p_1$ goes to 0 (or 1) is thus a reduction of the oscillation amplitude towards their initial value. 

These findings show how the most relevant case for our purposes is the case when $p_1^S=p_2^S=1/2$, when both entanglement and discord vanish and revivals occur.

\section{Correlations in a classical-quantum state}
We have shown that quantum correlations can revive also for a classical environment, and without back-action.
To look more closely at this, let us  consider a so-called classical-quantum state of the two qubits labeled $A$ and $B$, and the classical environments labeled $a$ and $b$. This state is initially given by
\begin{equation}
\rho_{QE}(0)=\rho_Q^{AB}(0)\otimes\sum_{i=1}^2 p_i |i\rangle_{EE}^{a~a}\langle i|\otimes\sum_{j=1}^2 p_j |j\rangle_{EE}^{b~b}\langle j|,
\end{equation}
where $\rho_Q^{AB}$ is the quantum state of the two qubits, and $|1\rangle_E ^a$ and $|2\rangle_E^a$ are two ``classical" basis states that environment $a$ can take, and $|1\rangle_E ^b$ and $|2\rangle_E^b$ are classical basis states of environment $b$. The classical environments can only be in separable mixtures of their basis states; in our case, $p_1=p_2=1/2$. The single-qubit map on qubit $A$ is now equivalent to applying a unitary transform
\begin{equation}
U^{Aa}(t)=\sum_{i=1}^2U_i^A(t)\otimes|i\rangle_{EE}^{a~a}\langle i|
\end{equation}
to qubit $A$ and its environment $a$, where $U_i^A$ is defined in Eq. (\ref{eq:singlemap}) for $S=A$, and tracing over the environment. The environment acts as ``control system" for what unitary operation is applied to the qubit.
The map $U^{Bb}(t)$ for qubit $B$ is constructed analogously. During time evolution according to $U^{Aa}$ and $U^{Bb}$, the states of the classical environments do not change. The qubits obviously do not affect their environments, and back-action by the environments on the qubits is clearly not present. The two-qubit map in Eq. (\ref{globalrandomfieldmap}) can be obtained by applying $U^{Aa}(t)\otimes U^{Bb}(t)$ to both qubits and environments, and tracing over the latter.

We can now quantify the correlations in the classical-quantum system exactly as for a quantum-only system, using the definitions in Eq. (\ref{quantifiers}). Since the subsystems $Aa$ and $Bb$ both evolve under local unitary operations, the quantum and classical correlations between $Aa$ and $Bb$ are obviously constant.
If we trace over the environments, then the correlations that remain between $A$ and $B$ were found to oscillate. If we trace over the qubits $A$ and $B$, then the environments $a$ and $b$ remain completely uncorrelated throughout the evolution.

If we look at the qubit-environment state $Aa$, ``entanglement" arises neither within this state, nor in $Bb$. Interestingly, ``quantum discord" may arise in the classical-quantum state $Aa$, and in $Bb$, however, one easily realizes that the presence of ``quantum discord" is not necessary for the revival of quantum correlations between the two qubits~\footnote{We have enclosed ``entanglement" and ``quantum discord" in quotes, since the environment is really a classical state.}.
The classical environments cannot store any quantum correlations on their own, and they do not become entangled with their respective quantum systems. But through the classical record they keep memory of what unitary operation has been applied to the qubits, they can obviously nevertheless play a vital part in reviving the quantum correlations, in particular the entanglement, that was initially present in the quantum systems $A$ and $B$. This remains the case even without back-action.
Although the environments can hold information about the quantum systems, it may be somewhat misleading to talk about ``information flow from the system(s) to the environment(s)", or ``information backflow from the environment(s) to the system(s)". This is because the information an environment holds about a system is due to what action the environment has on the system, and not the other way around.

After this explanation, it may seem trivial that classical environments can store quantum correlations in the sense that the state $Aa-Bb$ can be entangled, but the state $A-B$ when the classical environments are traced out is less entangled or even separable. This fact has nevertheless been overlooked when explaining the mechanisms for revivals of quantum correlations.
Although in our case the environments are classical, quantum correlations in the system can obviously revive when there is no back-action also for quantum environments.  The considerations above can also easily be generalized to arbitrary number of dimensions for  both systems and environments.

\section{A class of global time evolutions giving no back-action}
It is now clear that classical and quantum correlations may revive even when environments are modelled as classical, and for no back-action, and why this happens. There is no back-action when the system does not affect itself via the environment. This must be the case when the environment dynamics is unaffected by the system. We will now discuss a class of global (system-environment) evolutions where back-action is absent. We show that for this class of time evolutions, any reduced dynamics of the system that can be obtained with a quantum environment, can also be obtained with a classical environment.

We consider time evolutions (in the finite-dimensional case) described by completely positive (CP) maps \cite{nielsenchuang}. Such time evolution of a quantum system $S$ can be described as a unitary transform acting on $S$ and some environment $E$, followed by tracing over $E$. This will yield a CP map for the time evolution of $S$,
\begin{eqnarray}
\label{eq:kraus}
\rho_S(t)&=&{\rm Tr}_E[U(t)\rho_S(0)\otimes \rho_E(0)U^\dagger(t)]\nonumber\\
&=&\sum_k A_k(t)\rho_Q(0)A_k^\dagger(t).
\end{eqnarray}
Here $A_k(t)$ are the Kraus operators for the CP map, satisfying $\sum_k A_k^\dagger(t)A_k(t)=\openone_S$, and $\rho_E(0)$ is some initial state of the environment. (Given a CP map, we can construct $U(t)$ so that $\rho_E(0)=|0\rangle_{EE}\langle 0|$.)

The unitary transformation $U(t)$ will in general also affect the state of the environment so that not all completely positive maps on a quantum system can be described trough unitary evolution of the system plus an unchanged environment. Denote the basis states for $S$ and $E$ by $\ket{i}_S$ and $\ket{j}_E$, respectively. We also assume that a classical environment could be prepared in any mixture of its basis states, and a quantum environment in any linear superposition of these states. We moreover let the environment undergo an assigned evolution $U_E(t)$ which is unaffected by the system, meaning that its evolution is just $\rho_E(t)=U_E(t)\rho_E(0)U_E^\dag(t)$. One realizes that the condition
\begin{equation}
\label{eq:physical request}
_S\bra{i} _E\bra{j}U_E^\dag(t) U(t) \ket{l}_E \ket{k}_S = 0 \; \mathrm{unless}\; j=l
\end{equation}
is necessary for the state of the environment to be unaffected by the system. It is also possible to show that this condition is sufficient for the environment not to be affected by the system provided that it is initially prepared in a classical mixture of the chosen basis states. This can be shown by looking at $\rho_E(t)$ and, using the condition in Eq.~(\ref{eq:physical request}), requiring that it satisfies the equality $\rho_E(t)=U_E(t)\rho_E(0)U_E^\dag(t)$. If Eq.~(\ref{eq:physical request}) is satisfied, then $U(t)$ must be of the form
\begin{equation}
\label{eq:Unoback}
U(t) = \sum_{j} U_j^S(t)\otimes U_E(t)\ket{j}_{EE}\bra{j},
\end{equation}
where $U_j^S(t)\equiv\ _E\bra{j}U_E^\dag(t)U(t)\ket{j}_E$ are arbitrary unitary transforms on the system.
The time evolution of the quantum system can therefore be described by the random application of unitary transformations,
\begin{equation}\label{RandomExternalFieldMap}
\rho_S(t) = \sum_{j} p_{j} U_{j}^S(t)\rho_S(0) U_{j}^{S\dagger},
\end{equation}
where the time-independent probabilities $p_{j}$ are determined by the initial state of the environment. This state can be chosen as a quantum state $\ket{\psi}_E = \sum_{j} \sqrt{p_{j}}|j\rangle_E$, but the ``classical" state $\rho_E = \sum_{j} p_{j} |j\rangle_{EE}\langle j|$ will yield exactly the same CP map for the system. Importantly, therefore, if the global time evolution has the form of Eq.~(\ref{eq:Unoback}), there is no loss of generality in modelling the environment as classical. It is worth noticing that the map acting on the system given in Eq.~(\ref{RandomExternalFieldMap}) is exactly of the ``random external fields'' type. The map considered in our physical model in Sec.~\ref{model} is of this form.

We remark that if the environment is in a quantum superposition state of the relevant basis states, then the reduced state of the environment can be affected by the system, since in this case the system and environment may become entangled. More specifically, diagonal elements of the environment reduced density matrix evolve unaffected by the system, according to $U_E(t)$ only. Non-diagonal elements, however, including their absolute values, can be affected by the system dynamics. Nevertheless, this behavior in the environment reduced density matrix has no effect on the reduced dynamics of the system.

\section{Revivals and non-Markovianity}
It also looks appealing to interpret the revival of correlations by referring to intrinsic characteristics of two-qubit system evolution itself. Revival of correlations can be connected to the degree of non-Markovianity of the map, which effectively is the memory that the state of the system has of the correlations it owned in the past. In our model of the two-qubit dynamics, negative rates in the corresponding single-qubit master equation appear, as occurs e.g. for a phase noisy laser \cite{cresser2010OptComm}, of which our model is a special case. Negative rates typically indicate non-Markovianity of the map \cite{andersson2010Arxiv}. Both the single-qubit and the two-qubit time evolution in our model is thus non-Markovian. In particular, in Fig.~\ref{fig:entanglement} we see that the state $\rho(t_1)$ at $t_1=\pi/4g$ has both zero entanglement and zero discord. Nevertheless, both entanglement and discord subsequently revive. If we take the intermediate state $\rho(t_1)$ as initial state, then since correlations cannot exceed their initial values as shown by Eq.~(\ref{DCinequality}), this state remains non-entangled and has zero discord. Therefore,
$\Lambda(t,t_1)\rho(t_1)\neq \Lambda(t,t_0)\rho(t_0)$ where $\rho(t_1)=\Lambda(t_1,t_0)\rho(t_0)$. Our map does not satisfy the Markovian composition law $\Lambda(t,t_0)=\Lambda(t,t_1)\Lambda(t_1,t_0)$ for completely positive maps~\cite{alickibook}, where $\Lambda(t_2,t_1)$ has been defined in Eq.~(\ref{globalrandomfieldmap}). This means that the time evolution from the intermediate state $\rho(t_1)$ onwards is not a CP evolution.

One can render the association between revival of correlations and the non-Markovianity of the time evolution quantitative as follows. We generalize the quantifier of non-Markovianity introduced in Ref.~\cite{plenio2010PRL}, which is based on the increase of entanglement between the system and an isolated ancilla, to the time dependent form
\begin{equation}
\label{eq:nonmarkovquant}
\mathcal{I}^{E}(t)=\int_{t_0}^{t}\left|\frac{d E[\rho_{AB}(t')] }{d t'}\right|dt'-\Delta E(t),
\end{equation}
where $E$ is defined in Eq.~(\ref{quantifiers}) and $\Delta E(t)=E(t)-E(t_0)$, with $t_0\le t$. Eq.~(\ref{eq:nonmarkovquant}) can be applied to the single-qubit map (\ref{eq:singlemap}) in our model by initially preparing qubit $B$ (system) and another qubit $A$ (ancilla) in a maximally entangled state, for example $\rho_{AB}(0)=\ket{2_+}\bra{2_+}$. The evolved state is then given by $\rho_{AB}(t)= \cos^2 (g t)\ket{2_+}\bra{2_+}+\sin^2 (g t)\ket{1_-}\bra{1_-}$ while the closest separable state $\sigma_{\rho_{AB}(t)}= \frac{1}{2}[\ket{2_+}\bra{2_+}+\ket{1_-}\bra{1_-}]$ is time independent. We find that $\mathcal{I}^{E}(t)$ monotonically increases in the same time regions when revivals of entanglement and quantum discord occur, as shown in Fig.~\ref{fig:entanglement}. As a consequence of this behavior, $\mathcal{I}^{E}(t)$ would increase to infinity in the ideal case of zero dissipation. In a more realistic scenario, the periodic behavior of various quantifiers will be limited to a small number of periods and  $\mathcal{I}^{E}(t)$ will increase until dissipation destroys the revival of correlations.

The entanglement in Eq. (\ref{eq:nonmarkovquant}) is that between a single qubit subject to some single-qubit time evolution, and an ancilla qubit which does not evolve. One can also define a corresponding quantifier for which the ancilla qubit is also subject the same single-qubit time evolution as the first qubit. The entanglement in this quantifier would then correspond exactly to that between the two qubits in our model, as plotted in Fig.~\ref{fig:entanglement}. This makes the connection between the revival of two-qubit system correlations and the non-Markovianity.

\section{Conclusions}
We have considered a simple model where random external classical fields act locally on two independent qubits. We found that collapses and revivals of entanglement between the qubits, as well as decreases and increases of quantum discord and classical correlations, occur even when the environment is unaffected by the system qubits. This is in contrast to the common interpretation that revival of correlations is caused by transfer of back-action-induced correlations between system and environment \cite{lopez2008PRL,Lopez2010PRA}.
In terms of correlations in a classical-quantum state, we have explained why quantum correlations can revive also for classical independent environments on which back-action is not present.
We have shown that the increase of correlations in our system can be connected with the increase of a parameter used to quantify non-Markovianity of the dynamics of the individual qubits. If randomness would be introduced in the amplitude instead of in the phase, as considered in this paper, one may expect that revivals of correlations should also occur.

Furthermore, we have discussed a class of global time evolutions for which back-action is absent, even if the reduced dynamics of the environment can be affected by the system. We have shown that for this class of time evolutions, any system dynamics obtained with a quantum environment can also be obtained by modelling the environment as classical. This global time evolution results in a reduced dynamics for the system precisely of the form that arises from random external fields. Suppose that a quantum system plus its environment undergo unitary evolution, so that there is no additional environment. It is natural to think of a classical environment as a ``macroscopic" system that should not be significantly affected by the ``microscopic" quantum system it interacts with. Interestingly, what we found means that conversely, if a quantum system either does not affect its environment or influences the latter in a way that does not result in back-action, then there is no loss of generality in modelling the environment as classical, as far as the possible time evolutions of the system are concerned.

\begin{acknowledgments}
Some of the authors, B.B. and G.C., wish to acknowledge discussions with Prof. R. Alicki. E.A. gratefully acknowledges partial support from EP/G009821/1.
 \end{acknowledgments}

\end{document}